\documentclass[12pt]{article}
\usepackage{amsmath, amssymb}
\usepackage{float}
\usepackage{geometry}
\usepackage{hyperref}
\hypersetup{colorlinks=true, linkcolor=blue, citecolor=blue, urlcolor=blue}
\usepackage{graphicx}
\usepackage{subfig}
\usepackage[numbers,sort&compress]{natbib}

\title{\bf Thick brane in Palatini formalism with a non-minimally coupled bulk scalar field }

\author{
\bf Tahereh Azizi\thanks{Email: t.azizi@umz.ac.ir} \and
\bf Mojtaba Alimoradi
}

\date{
\it Department of Theoretical Physics, Faculty of Sciences,\\
	University of Mazandaran,\\
		P. O. Box 47416-95447, Babolsar, IRAN\\
}
\begin{document}
	\maketitle
\begin{abstract}
	We study a thick brane scenario within the Palatini formulation of gravity, where the metric and affine connection are treated as independent variables. By introducing a non-minimal coupling between a bulk scalar field and the Ricci scalar, we obtain analytic solutions under a flat, four-dimensional Poincaré-invariant metric with a kink-like scalar configuration. The warp factor exhibits a bell-shaped profile, while the scalar potential forms a symmetric volcano-like structure, characteristic of a finite-thickness brane. The corresponding energy density is regular and localized, featuring a central peak with symmetrically placed negative minima. 	
	Through the analysis of linear tensor perturbations, we derive a Schrödinger-like equation with supersymmetric factorization, ensuring the absence of tachyonic modes and thus the stability of the background configuration. The effective potential also takes a volcano-like form that supports a localized graviton zero mode, confirming the recovery of four-dimensional gravity on the brane. A numerical study of the massive Kaluza--Klein spectrum reveals the progressive delocalization of massive modes into the bulk. 	
	
	Our results demonstrate a stable and physically consistent thick brane configuration within the Palatini gravity framework, offering new insights into gravity localization and braneworld phenomenology.
\end{abstract}

%
%

	\section{Introduction}\label{sec1}
	
The hypothesis that our observable universe might constitute a four-dimensional brane embedded within a higher-dimensional bulk space has attracted considerable attention in recent decades \cite{Akama:1982jy,Rubakov:1983bb,Visser:1985qm,Antoniadis:1990ew,Gremm:2000dj}. Such braneworld scenarios emerge naturally in string theory and other fundamental frameworks, providing compelling approaches to several persistent challenges in modern physics. These include resolving the hierarchy problem, elucidating the nature of dark energy, explaining inflationary cosmology, and understanding the localization mechanism for matter fields \cite{Arkani-Hamed:1998jmv,Dvali:2000hr,Maartens:2003tw,Brax:2004xh,Koyama:2004ap,Tsujikawa:2004dm,Nozari:2007qg,Nozari:2009ys,Neupane:2010ey,Guo:2011qt,Azizi:2011ys,Doolin:2012ss,Neupane:2014vwa,Rani:2016wpa,Haghshenas:2022gka,Jalalzadeh:2023upb,Guo:2023mki,Zhou:2025dyr}.
	
Initial efforts to integrate extra dimensions, exemplified by Kaluza-Klein (KK) theory, primarily investigated compactified dimensions \cite{Overduin:1997sri}. These pioneering works established the foundation for higher-dimensional unification by geometrically interpreting additional fields as manifestations of extra-dimensional components. Nevertheless, such frameworks face inherent limitations stemming from their compact nature and constrained localization properties. A major breakthrough came with the introduction of the Randall-Sundrum (RS) models, which allow for non-compact extra dimensions and employ a warped geometry to localize gravity on a brane. In the RS1 model, the exponential warp factor addresses the hierarchy problem \cite{Randall:1999ee}, while the RS2 model demonstrates that even with a single brane and an infinite extra dimension, four-dimensional gravity can be effectively recovered on the brane \cite{Randall:1999vf}. These scenarios, while insightful, typically assume an idealized thin brane with delta-function structure, which may not capture the full complexity of brane formation or internal dynamics.

In more realistic field-theoretic models inspired by fundamental theories, it is expected that a minimum length scale exists, suggesting that branes should possess finite thickness rather than being idealized as infinitesimally thin surfaces. This realization has motivated the development of \textit{thick brane} models, in which branes arise dynamically from scalar field configurations rather than being inserted artificially into the action. Such models eliminate the singularities associated with thin branes and allow for richer gravitational and field-theoretic behavior. In these setups, the spectrum of gravitational fluctuations includes a normalizable massless zero mode responsible for standard four-dimensional gravity, as well as a tower of massive KK excitations that generate small-scale corrections to Newtonian gravity. Moreover, in such scenarios, the scalar fields do not merely act as passive bulk fields; instead, they constitute the very substance of the brane, giving rise to a smooth and regular geometry \cite{Gremm:1999pj,Csaki:2000fc,Kobayashi:2001jd,Bronnikov:2003gg,Barbosa-Cendejas:2006cic,Liu:2007ku,Dzhunushaliev:2009va,Liang:2009zzc}.

A particularly interesting aspect is that branes can form naturally, without explicit introduction in the action. Furthermore, in thick brane scenarios, gravity is often coupled to background scalar fields, which provide the structural foundation of the brane itself. While extensive research has examined thick branes with minimal coupling, investigations of non-minimally coupled scalar fields remain relatively scarce, despite their potential to uncover a wider range of physical phenomena \cite{Farakos:2005hz,Pasipoularides:2007zz,Andrianov:2007tf,Guo:2011wr,Liu:2012gv,Rohman:2023apv}. The non-minimal coupling between the scalar field and the Ricci scalar introduces additional degrees of freedom and modifies the dynamics of both the scalar field and the geometry. This coupling affects the structure of the brane, the localization properties of the gravitational modes, and the form of the KK spectrum. In addition to scenarios involving scalar fields, thick brane models have been extensively investigated within the context of modified theories of gravity \cite{Bazeia:2008zx,Zhong:2010ae,Liu:2011wi,Xu:2014jda,Bazeia:2015owa,Zhong:2015pta,Rosa:2020uli,Lobao:2022inw,Peyravi:2022ubf,Almeida:2024elf,Guo:2024izl,Moreira:2024unj,Sorkhi:2025rfc}. 

The aim of this paper is to investigate thick brane solutions in Palatini gravity with non-minimally coupled bulk scalar fields. The Palatini formalism, which treats the metric and the affine connection as independent variables, offers a powerful alternative to the traditional metric formalism for studying gravitational systems. In standard General Relativity and minimally coupled scenarios, both formalisms yield equivalent field equations. However, in theories with non-minimal coupling between scalar fields and curvature—such as those found in many extended gravity models—the Palatini and metric formalisms lead to fundamentally different dynamics. As a result, new gravitational phenomena can emerge in four-dimensional theories formulated in the Palatini framework \cite{Bauer:2008zj,Olmo:2011uz,Wang:2012rva,Fan:2015rha,Kozak:2018vlp,Wang:2018akv}, as well as in five-dimensional $f(R)$ gravity, which differ from those derived using the metric formulation of the same theories \cite{Gu:2014ssa,Bazeia:2014poa,Gu:2018lub}.
This distinction is particularly important in the context of thick brane models, as non-minimal coupling affects not only the structure and stability of the brane but also the localization of gravitational modes and corrections to four-dimensional gravity. The independence of the connection in the Palatini approach allows for a broader class of geometrical solutions, which can significantly influence the resulting brane configuration and the effective physics on the brane.
Our model incorporates a non-minimally coupled scalar field in the bulk, which plays a central role in stabilizing the brane and shaping the effective potential that governs graviton localization. We derive the field equations from a generalized action that includes contributions from the scalar field, its potential and the coupling function. We then solve the resulting system to construct thick brane solutions and analyze the corresponding graviton wavefunctions. Our results demonstrate that the zero mode is sharply localized around the brane core, ensuring recovery of standard gravity at large distances.
	
The organization of this paper is as follows: in Sec.~\ref{sec2}, we obtain the equations of motions for the non-minimally coupled bulk scalar field with the Palatini Ricci scalar and construct the braneworld scenario by considering a warp geometry in the five-dimensional of space-time. In Sec.~\ref{sec3}, we analyze the linear stability of the thick brane system under tensor perturbations, investigating both the localization of the graviton zero mode (ensuring 4D gravity recovery) and the massive Kaluza-Klein modes. The conclusion is presented in Sec.~\ref{sec4}.
	
	\section{ The model and the solution}\label{sec2}
We consider a five-dimensional gravitational action in the Palatini formalism, which includes a non-minimal coupling between bulk scalar fields and the Ricci scalar
\begin{equation}\label{action1}
	S=\int d^5x \sqrt{-g} \left[f(\phi)\hat{R} +\mathcal{L_{\phi}}\right],
\end{equation}
	where $g$ is the determinant of the spacetime metric $g_{\mu\nu}$ and $\hat{R}$ refers to the five dimensional Ricci scalar, , derived by contracting the Ricci tensor $\hat{R}_{\mu\nu}$. This tensor is constructed from the independent connection $\hat{\Gamma}^\sigma_{\mu\nu}$ as
	 	\begin{equation}\nonumber
	 		\hat{R}_{\mu\nu}=\hat{\Gamma}^\sigma_{\nu\mu,\sigma}-\hat{\Gamma}^\sigma_{\sigma \mu,\nu}+\hat{\Gamma}^\sigma_{\sigma \rho}\hat{\Gamma}^\rho_{\nu \mu}-\hat{\Gamma}^\sigma_{\nu\rho}\hat{\Gamma}^\rho_{\sigma \mu}\,. 
	 		\end{equation}
	In addition, $\mathcal{L_{\phi}}$ is the Lagrangian of the scalar field, which is defined as 
			\begin{equation}\label{phi-lagrangy}
		\mathcal{L_{\phi}}=-\frac{1}{2}	\nabla^{\mu}\phi \nabla_{\mu}\phi-V(\phi),
		\end{equation}
		
where $V(\phi)$ is the potential of the scalar field. It should be noted that in Palatini models, the metric and connection are treated as independent varaibles. If a minimal coupling between the scalar field and the Ricci scalar is present, the resulting field equations for both will be identical in form. Conversely, when non-minimal coupling exists between the scalar field and the Ricci scalar, distinct field equations emerge, reflecting the complexities introduced by this coupling. In this paper, Greek indices $\mu, \nu, \ldots$ denote the five-dimensional coordinates, while Latin indices $i, j, \ldots$ refer to the four-dimensional coordinates. The extra dimension $x^{4}$ is denoted by $y$. Varing action (\ref{action1}) with respect to the metric, we obtain the following field equation 
\begin{equation}
f(\phi)\left(\hat{R}_{\mu\nu}-\frac{1}{2} \hat{R} g_{\mu\nu}\right)=T_{\mu\nu}^{(\phi)}, \label{eq:metric}
\end{equation}

where $T_{\mu\nu}^{(\phi)}$ is the energy momentum tensor of the scalar field and is defined as
\begin{equation}
T_{\mu\nu}^{(\phi)}=\nabla_{\mu}\phi\nabla_{\nu}\phi-g_{\mu\nu}\left[\frac{1}{2}\nabla_{\sigma}\phi\nabla^{\sigma}\phi+V(\phi)\right], \label{t-phi}
\end{equation}
where $\nabla_{\mu}$ is a covariant derivative associated with the usual Levi-Civita connection. Contracting Equ (\ref{eq:metric}), we get 
\begin{equation}\nonumber
	\frac{-3}{2}f(\phi)\hat{R}=T,
\end{equation}
where $T$ is the trace of the energy-momentum tensor of the scalar field. This relation implies that there is an algebric relation between the gravity and source field. As a result, $\hat{R}$ can be expressed by $T$. 
Variation of the action (\ref{action1}) with respect to the connection yields
\begin{equation}
	\hat{\nabla}_{\sigma} \left(\sqrt{-g} f(\phi) g^{\mu\nu}\right)=0. \label{eq:connection}
\end{equation}

Note that $\hat{\nabla}$ is the covariant derivative defined with respect to the independent connection $\hat{\Gamma}$ and is not compatible with the metric. This implies that the covariant derivative of the metric does not vanish ($\hat{\nabla}_{\sigma} g_{\mu\nu}\neq0$). 
Furthermore, varying the action with respect to the scalar field gives its equation of motion as

\begin{equation}
	\nabla^{2}\phi-\frac{dV}{d\phi}+\hat{R}\frac{df(\phi)}{d\phi}=0. \label{motion-phi}
\end{equation}

It should be noted that the field equations (\ref{eq:metric}) are different from the field equations in non-minimal coupling between bulk scalar field and Ricci scalar in metric formalism \cite{Pasipoularides:2007zz}. As it is conventional in Palatini formalism, in order to solving equation (\ref{eq:metric}) and (\ref{eq:connection}), we introduce an auxiliary metric in the following form
\begin{equation}\label{auxilar metric}
\sqrt{-q}q^{\mu\nu}\equiv\sqrt{-g}f(\phi)g^{\mu\nu},
\end{equation}

where $\hat{\nabla}_{\sigma}(\sqrt{-q} q^{\mu\nu})=0$. From Equations  (\ref{eq:connection}) and (\ref{auxilar metric}), it follows that 
\begin{equation}\label{auxilar metric2}
q^{\mu\nu}=f(\phi)^{-2/3} g^{\mu\nu};\;\;\; q_{\mu\nu}=f(\phi)^{2/3} g_{\mu\nu},
\end{equation}

indicating that the two metrics are conformally related. It is important to note that  the conformal relationship between $q_{\mu\nu}$ and $g_{\mu\nu}$ is governed exclusively by the scalar-field-dependent non-minimal coupling function. The independent connection $\hat{\Gamma}$ can be expressed as the Levi-Civita connection of the new metric $q_{\mu\nu}$:
\begin{equation}\label{auxilar metric3}
\hat{\Gamma}^\sigma_{\mu\nu}=\frac{1}{2}q^{\sigma\rho}\left(q_{\mu\rho,\nu}+q_{\nu\rho,\mu}-q_{\mu\nu,\rho}\right).
\end{equation}

Using the above equation, we can express the Palatini Ricci tensor and Ricci scalar in terms of their metric counterparts as
\begin{equation}\label{richi-pala}
\hat{R}_{\mu\nu}=R_{\mu\nu}(g)-\frac{1}{3f(\phi)}\left[3\nabla_{\mu}\nabla_{\nu}f(\phi)+g_{\mu\nu}\nabla_{\sigma}\nabla^{\sigma}f(\phi)\right]+\frac{4}{3f(\phi)^2}\nabla_{\mu}f(\phi)\nabla_{\nu}f(\phi)
\end{equation}
and 
\begin{equation}\label{r-pala}
	\hat{R}=R(g)-\frac{8}{3f(\phi)}\nabla_{\mu}\nabla_{\nu}f(\phi)+\frac{4}{3f(\phi)^2}\nabla_{\sigma}f(\phi)\nabla^{\sigma}f(\phi),
\end{equation}	
where $R_{\mu\nu}(g)$ and $R(g)$ denote the Ricci tensor and Ricci scalar in the metric formalism, respectively. Now, Eq. (\ref{eq:metric}) can be recast using Eqs. (\ref{richi-pala}) and (\ref{r-pala}) as
\begin{equation}
	G_{\mu\nu}=\frac{1}{f(\phi)}\left[ \nabla_{\mu}\nabla_{\nu}-g_{\mu\nu}\nabla_{\sigma}\nabla^{\sigma}\right]f(\phi)-\frac{4}{3f(\phi)^2}\left[\nabla_{\mu}f(\phi)\nabla_{\nu}f(\phi)-\frac{1}{2}g_{\mu\nu}\nabla_{\sigma}f(\phi)\nabla^{\sigma}f(\phi)\right]+\frac{T_{\mu\nu}^{(\phi)}}{f(\phi)}\,,
		\label{eq:field-equ1}
\end{equation}
where $G_{\mu\nu}$ is the Einstein's tensor. This demonstrates that Palatini gravity is dynamically equivalent to a metric theory with a modified matter source.
To derive the brane equations, we adopt a flat metric with four-dimensional Poincar$\acute{e}$ symmetry expressed as 
\begin{equation}
d s^2=a^2(y)\eta_{ij}d x^{i}d x^{j}+d y^{2},\label{background metric}
\end{equation}
where $\eta_{ij}$ represents the Minkowski metric and $a(y)$ is the warp factor, which is assumed to depend solely on the extra dimension. As is typical in standard braneworld scenarios, we assume that the scalar field depends only on the extra dimension. Therefore, by utilizing equations (\ref{t-phi}) and (\ref{background metric}), the components of the energy-momentum tensor of the scalar field are expressed as
\begin{align}
	T_{ij}&=-\frac{1}{2}a^2 \left(\phi'^2+2V\right)\eta_{ij},
	\label{e-m-phi1}\\
	T_{55}&=\frac{1}{2}\left(\phi'^2-V\right),
	\label{e-m-phi2}
\end{align}
where the prime denotes a derivative with resprct to the extra dimention. From equations (\ref{e-m-phi1}) and (\ref{e-m-phi2}), we obtain the trace of the energy momentum tensor as $T=-5V-\frac{3}{2}\phi'^2 $. Additionally, the scalar field's equation of motion is given by:
\begin{equation}
 \phi''+4\frac{a{'}}{a}\phi{'}-\frac{dV}{d\phi}+\left(\frac{10}{3}V+\phi'^{2}\right)\frac{d \ln{f(\phi)} }{d\phi}=0.\label{clien-gordon }	
\end{equation}

In developing our thick brane-world scenario, we must solve the coupled system of Eqs. (\ref{eq:field-equ1}) and (\ref{clien-gordon }). Conventional methods involving either superpotentials \cite{DeWolfe:1999cp} or ansatze for the potential \cite{Liu:2011wi} fail here because of the intricate structure of Eq. (\ref{eq:field-equ1})'s source terms. We therefore adopt the strategy of \cite{Gu:2014ssa}, working instead with Eqs. (\ref{eq:metric}) and (\ref{auxilar metric}) as our starting point. 
 Given the conformal relationship between $q_{\mu\nu}$ and $g_{\mu\nu}$, established in Eq. (\ref{auxilar metric}), it becomes advantageous to adopt the following ansatz for the auxiliary metric:
\begin{equation}
d\tilde{s}^2=q_{\mu\nu}dx^{\mu}dx^{\nu}=f(\phi)^{\frac{2}{3}}\left[a^2(y)\eta_{ij}dx^{i}dx^{j}+dy^{2}\right].
\	
\end{equation}
Therefore, the field equations (\ref{eq:metric}) can be expressed as
\begin{align}
\frac{u''}{u}+\frac{a'}{a}\frac{u'}{u}-2\frac{u'^{2}}{u^{2}}	&=-\frac{\phi'^{2}}{3f(\phi)},
	\label{field3}\\
\frac{u''}{u}+\frac{a'}{a}\frac{u'}{u}+2\frac{u'^{2}}{u^{2}}	&=-\frac{-2V}{3f(\phi)},
	\end{align} 
	where we have defined $u=a\,f(\phi)^{\frac{1}{3}}$. Here, the equation of motion for the scalar field (\ref{clien-gordon }) is inherently satisfied as a result of the Bianchi identity.
 It is important to note that the system can be significantly simplified by establishing an appropriate relationship between $u(y)$ and $a(y)$. To achieve this, we assume $u(y) = k\, a^n(y)$ with ($n \neq 0$). We further propose a quadratic non-minimal coupling function of the form: $f(\phi)=1+\xi\phi^2$ where $\xi$ is the dimensionless coupling constant.

\begin{figure*}[htb]
	\begin{center}
		\includegraphics[width=8.5cm,height=8.5cm]{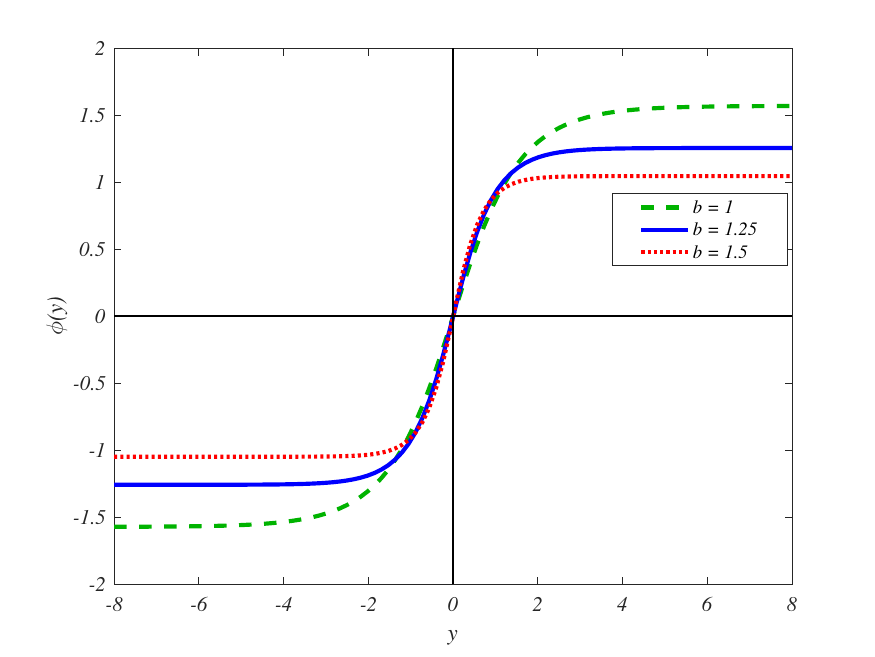}
			\end{center}
	\caption{The scalar field $\phi(y)$ versus $y$ for different
		values of $b$.}\label{phi-y}
\end{figure*}

\begin{figure*}[htb]
	\centering
	\subfloat[]{\includegraphics[width=0.48\linewidth]{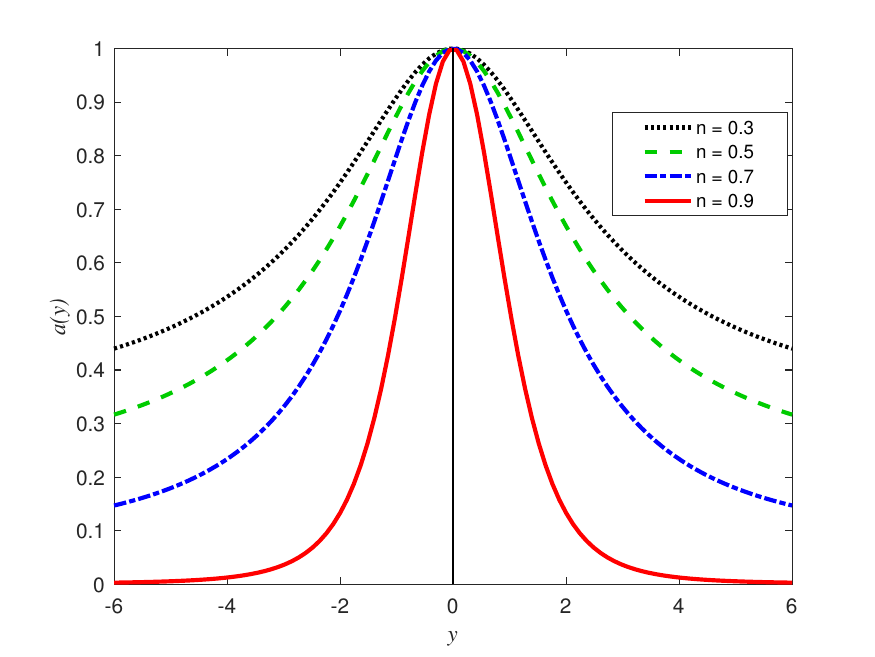}\label{fig:n}}
	\hfill
	\subfloat[]{\includegraphics[width=0.48\linewidth]{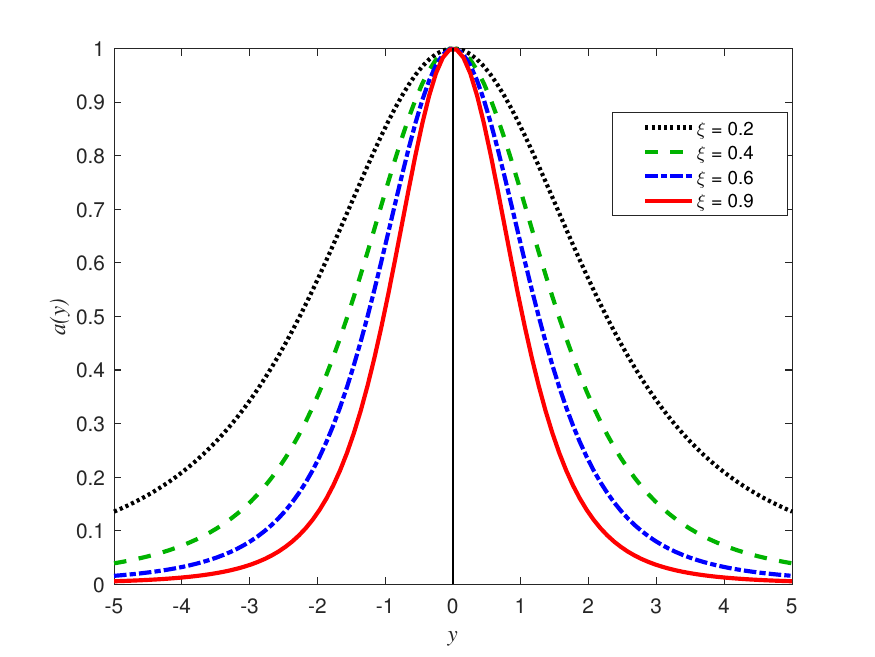}\label{fig:xi}}
	\caption{The shape of the warp factor associated to the function
		in Eq.(\ref{warp}) versus the extra dimension $y$ for various values of $n$ (a) and the coupling parameter $\xi$ (b). The parameters are set to $ b =c= 1$.}
	\label{fig-warp}
\end{figure*}

By requiring that the energy density approaches zero as $y \to \pm \infty$, the background scalar field must satisfy  the boundary conditions $\phi(y \to \pm \infty) < \infty$ and $\phi'(y \to \pm \infty) < \infty$. Consequently, the background scalar field can naturally be considered as a kink-like configuration. Following the approach of \cite{Bazeia:2014poa}, we adopt the analytical form
\begin{equation}
\phi(y) = \frac{1}{b} \arcsin[\tanh(b^2 y)]\,,
\label{scalar field-kink}
\end{equation}
where $b$ is a real parameter that determines the width of the warp factor. This choice allows for an explicit and smooth thick-brane solution while ensuring the desired asymptotic behavior of the energy density. The background scalar field is zero at $y=0$, and as $y\rightarrow\pm\infty$ the value of the background scalar field approaches the constant value ($\pm\pi/2b$). The behavior of the scalar field $\phi$ versus $y$ is depicted in figure \ref{phi-y}.
Using the relation (\ref{scalar field-kink}) for the scalar field, the analytical solutions of the warp factor is obtained as

\begin{equation}
a(y)=\left(\frac{1}{c^3}[1+\frac{\xi}{b^2}arc\sin(\tanh(b^2y))^2 ]\right)^{\frac{1}{3n-3}}\label{warp}.	
\end{equation}
In figure \ref{fig-warp}, we have plotted the warp function with respect to the extra dimension with varios values of the coupling parameter $\xi$ and $n$. As the figure shows, the warp function has the usual bell shape profileis and localized at $y=0$. Note that the brane is more localized for larger values of $\xi$ and $n$.  

\begin{figure*}[htb]
	\centering
	\subfloat[]{\includegraphics[width=0.48\linewidth]{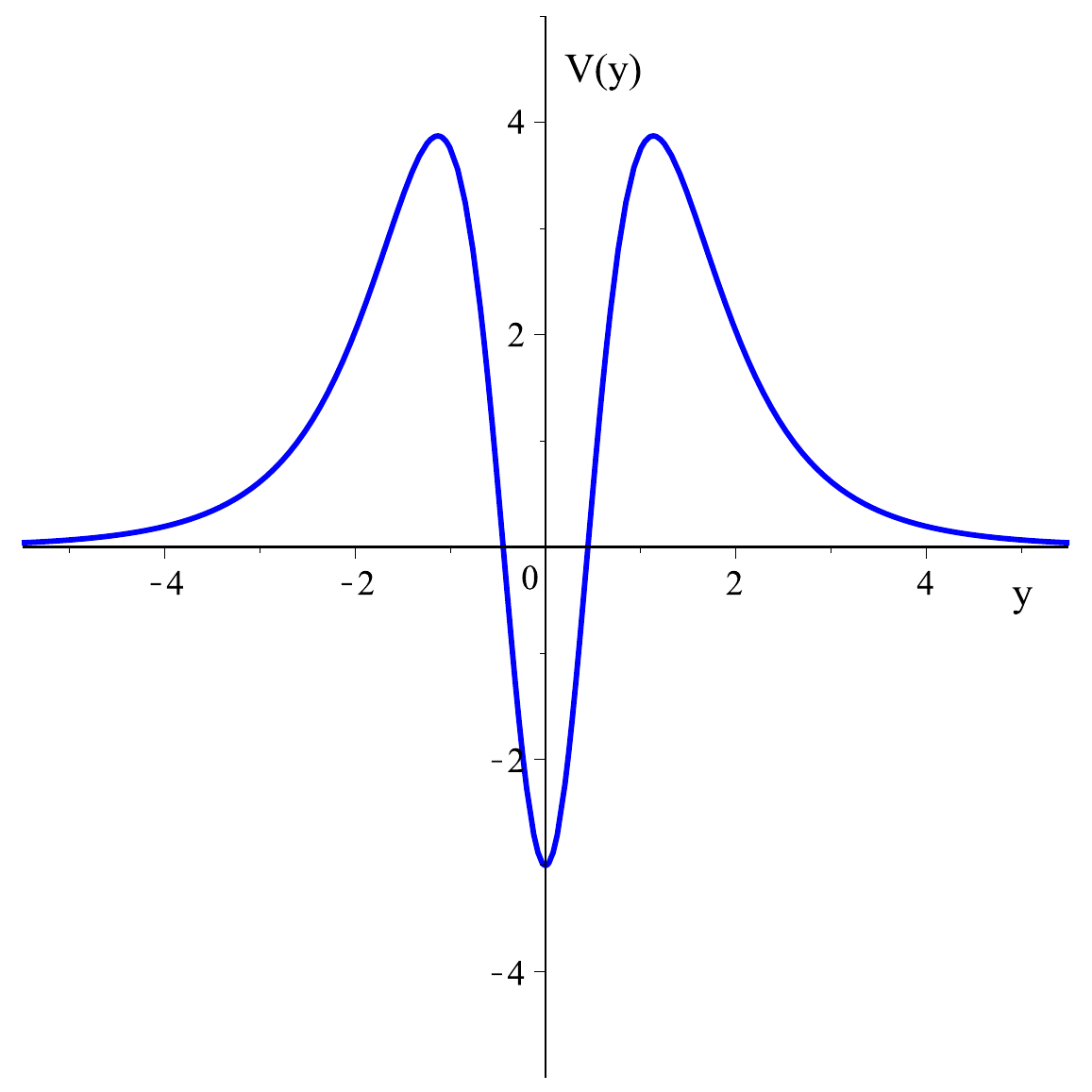}\label{fig-pot}}
	\hfill
	\subfloat[]{\includegraphics[width=0.48\linewidth]{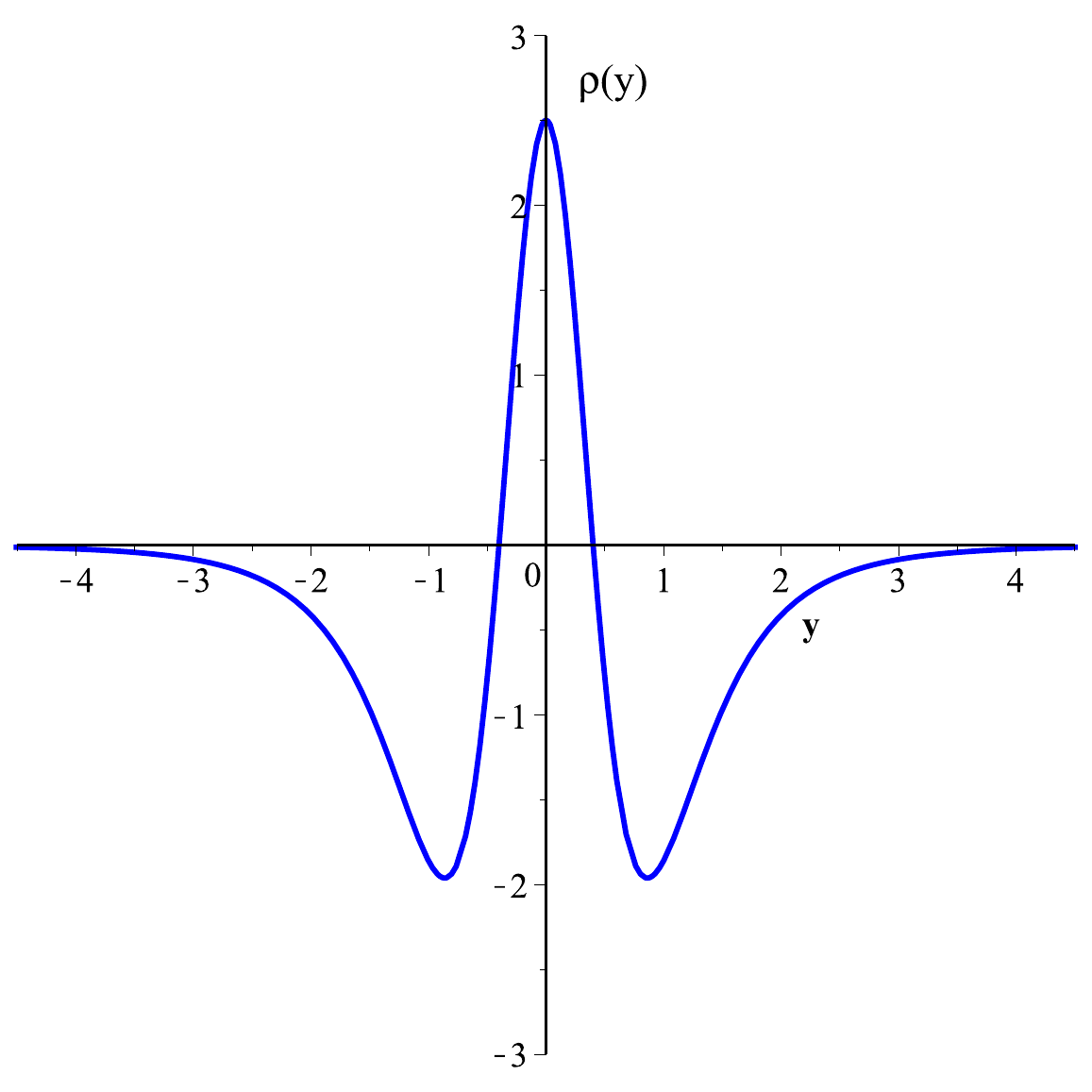}\label{fig-rho}}
	\caption{Potential of scalar field (a) and energy density (b) depicted as a function of extra dimension for $ b =c= 1$ and $n = \xi=0.9$.}
	\label{pot-rho}
\end{figure*}

Additionally, analytical expressions for the potential $V(y)$ and energy density $\rho = T_{00}$ have been obtained. However, due to their complexity, we omit the explicit forms here and instead illustrate their behavior in Fig.~\ref{pot-rho}. As shown in Fig.~\ref{fig-pot}, the potential $V(y)$ exhibits a $\mathbb{Z}_2$-symmetric volcano-like structure with \textit{maxima} at $y \approx \pm 2$ and a central \textit{minimum} at $y = 0$, characteristic of thick brane scenarios. The smooth profile suggests finite-width brane solutions, while the asymptotic behavior supports gravitational localization. Furthermore, Fig.~\ref{fig-rho} reveals the energy density distribution along the fifth dimension, where $\rho(y)$ peaks at $y = 0$ (localizing the brane), exhibits two negative minima, and vanishes asymptotically.

\begin{figure*}[htb]
	\begin{center}
				\includegraphics[width=8.5cm,height=8cm]{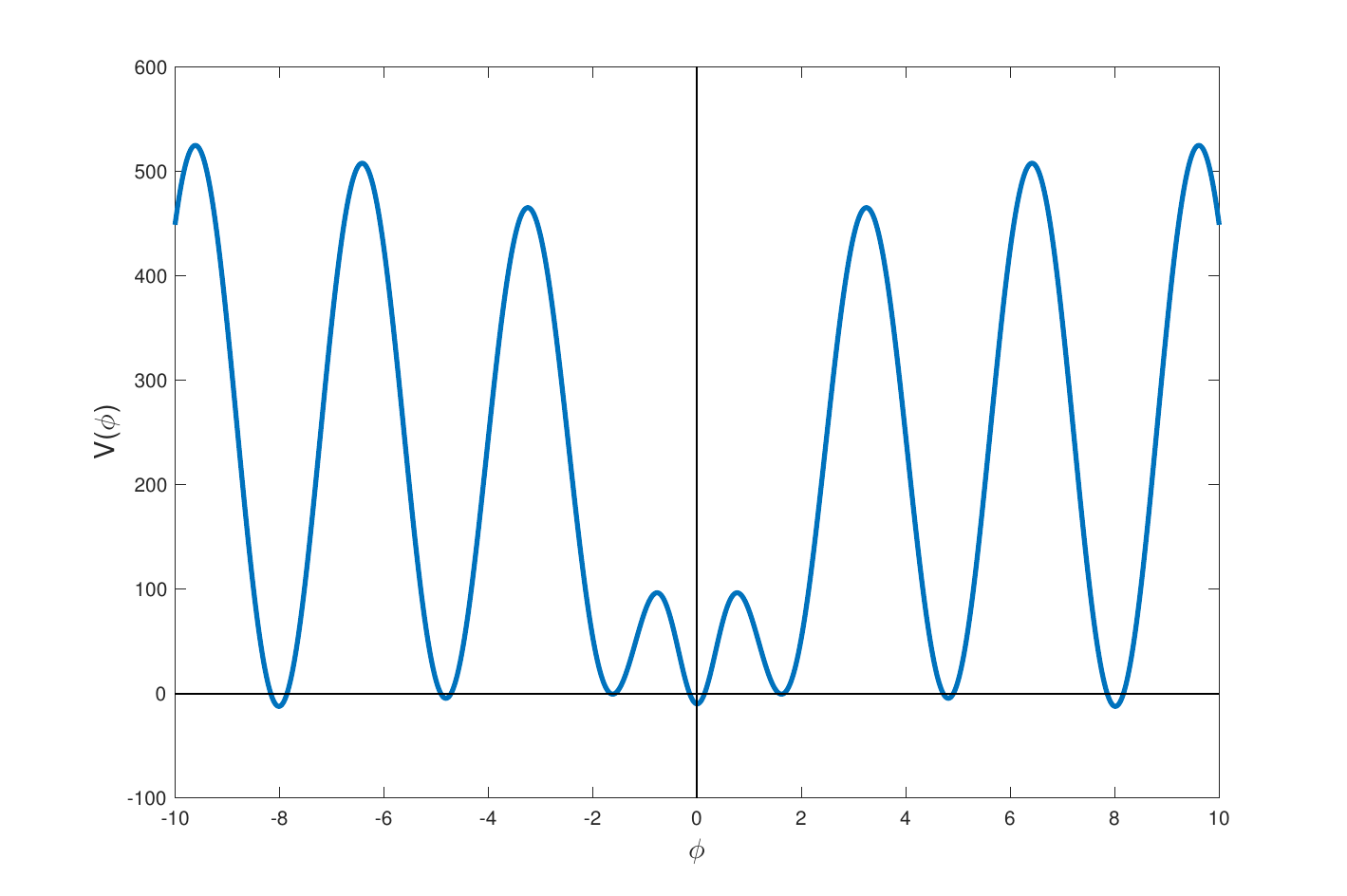}
	\end{center}
	\caption{The potential with respect to $\phi$. The numerical values are the same as figure \ref{pot-rho} }\label{figv-phi}
\end{figure*}

The potential for the scalar field versus $\phi$ has an exact solution as follows
\begin{equation}
V(\phi)	= -\frac{ b\phi\sin(2b\phi)}{n-1} + \frac{2\left( n + \frac{5}{3} \right) \xi (b\phi)^2 \cos^2(b\phi) + b^2 
	\cos^2(b\phi) (n-1)}{(\xi (b\phi)^2 + b^2)(n-1)^2} .
\end{equation}
 The scalar potential $V(\phi)$, depicted in Fig.~\ref{figv-phi}, exhibits a periodic oscillatory profile with a growing envelope as $|\phi|$ increases. The central region near $\phi = 0$ shows a deep minimum, indicating a natural localization point for the scalar field and the brane core. The growth of the potential at large field values ensures confinement of the scalar field, contributing to the overall stability of the thick brane.

	\section{Linear stability and the localization of gravity}\label{sec3}

The stability of thick brane solutions under small perturbations is crucial for their physical viability. In this section we investigate the linear stability of the braneworld by considering  small perturbations around the background metric
\begin{equation}
ds^2 = a^2(y)(\eta_{ij} + h_{ij})dx^i dx^j + dy^2,\label{metric-pertub}
\end{equation} 
where $h_{ij}=h_{ij}(x^i,y)$ reperesents the tensor fluctuation and for simplicity, we assume that it satisfies the transverese and traceless (TT) gauge conditions, namely $\partial^i h_{ij} = 0$ and $h=\eta^{ij}h_{ij}= 0$.
Thus, with the perturbated metric (\ref{metric-pertub}) and TT conditions, the perturbated $ij$-components of the Einstein tensor are obtained as
\begin{equation}
\delta G_{ij} = -\frac{1}{2} \square^{(4)} h_{ij} - \frac{1}{2}a^2  h''_{ij} - 2 aa' h'_{ij} + 3\left( a''a + a'^2 \right) h_{ij},\label{perturbed-G}
\end{equation}
where $\square^{(4)}\equiv\eta^{ij}\partial_{i}\partial_{j}$ is the 4D d'Alembertian. To perturb the field equations of the non-minimal Palatini brane model, we first rewrite equation (\ref{eq:field-equ1}) in the following form  
\begin{equation}
	G_{\mu\nu} = \frac{1}{f(\phi)}T_{\mu\nu}^{eff},\label{field-eff}
\end{equation}
where we have defined
\begin{equation}
	T_{\mu\nu}^{eff}=\left( \nabla_{\mu}\nabla_{\nu}-g_{\mu\nu}\nabla_{\sigma}\nabla^{\sigma}\right)f(\phi)-\frac{4}{3f(\phi)}\left[\nabla_{\mu}f(\phi)\nabla_{\nu}f(\phi)-\frac{1}{2}g_{\mu\nu}\nabla_{\sigma}f(\phi)\nabla^{\sigma}f(\phi)\right]+T_{ij}^{(\phi)}.
	\label{eff-T}
\end{equation}
So, the $ij$ components of the field equation (\ref{eq:field-equ1}) can be obtained by virtue of the metric ansatz (\ref{background metric})
\begin{equation}
	G_{ij} = \eta_{ij}\left(\frac{a^2}{f(\phi)}\right)\left[f''(\phi)+\frac{a'}{a}f'(\phi)-\frac{2(f'(\phi))^2}{3f(\phi)}+\frac{1}{2}\left(\phi'^2+2V(\phi)\right)\right],\label{field-ij}
\end{equation}
Using equations (\ref{metric-pertub}), (\ref{eff-T}) and (\ref{perturbed-G}), the perturabation of the $ij$-components of the field equation (\ref{field-eff}) can be obtained as
\begin{align}
-\frac{1}{2} \square^{(4)} h_{ij} - \frac{1}{2}a^2  h''_{ij} - 2 aa' h'_{ij} + 3\left( a''a + a'^2 \right) h_{ij}=\\\nonumber\frac{1}{f(\phi)}\left(f'(\phi)h'_{ij}+\left[-f''(\phi)-4\frac{a'}{a}f'(\phi)+\frac{2(f'(\phi))^2}{3f(\phi)}-\frac{1}{2}\left(\phi'^2+2V(\phi)\right)\right]h_{ij}\right)\,.\label{pertub1}
\end{align}

Multiplying the Eq.~(\ref{field-ij}) by $\frac{g^{ij}}{4}$  and combining the result with the above equation, we get
\begin{equation}
	\frac{1}{2} \square^{(4)} h_{ij} + \frac{1}{2}a^2  h''_{ij}+ 2 aa' h'_{ij}=\frac{\mathcal{F}}{2}\left(-h'_{ij}+	2\frac{a'}{a}h_{ij}\right),\label{pertub2}
\end{equation}
where $\mathcal{F} \equiv \frac{f'(\phi)}{f(\phi)} $ that encodes the effect of the non-minimal coupling between scalar field and Palatini Ricci scalar.
To see the behavior of $h_{ij}$, we transform this equation into the Schr\"{o}dinger-like equation. To simplify the analysis, we introduce the conformal coordinate $ z $ defined by
\begin{equation}
	\frac{dz}{dy} = \frac{1}{a(y)} \quad \Rightarrow \quad dy = a(y)\, dz\,.
\end{equation}

Consequently, when expressed in terms of the coordinate 
$z$, the background metric given in Equation (\ref{background metric}) takes the following form
\begin{equation}
	ds^2 = a^2(z) \left( \eta_{ij} dx^\mu dx^\nu + dz^2 \right)\,
\end{equation}
which is manifestly conformally flat. As a result, the linearized Einstein equations for the perturbations $h_{ij}(x,z)$, take the form:
\begin{equation}
	\left[ \partial_z^2 + 3 \frac{a'}{a} \partial_z + \Box^{(4)} \right] h_{ij}(x,z)
	= \mathcal{F} \left( -\partial_z h_{ij} + 2 \frac{a'}{a} h_{ij}, \right)\label{pertub-z}
\end{equation}
where hereafter we define $' \equiv \frac{d}{dz} $.  Now, we use the KK decomposition $h_{\mu\nu}(x,z) = e^{i p \cdot x} \tilde{h}_{ij}(z)$
where $ p^2 = -m^2 $ with $m$ representing the four-dimensional graviton mass. Substituting into the equation (\ref{pertub-z}) yields
\begin{equation}
	\tilde{h}''_{ij}(z) + \left( 3 \frac{a'}{a} +\mathcal{F} \right) \tilde{h}'_{ij}(z)
	- \left( 2\mathcal{F} \frac{a'}{a} + m^2 \right) \tilde{h}_{ij}(z) = 0\,.
	\label{eq:secondorder}
\end{equation}

To transform the above equation into the Schr\"{o}dinger-like form, we  eliminate the first derivative term by introducing a rescaled wavefunction $\tilde{h}_{ij}(z) = \Omega(z) \psi(z)$, where $ \Omega(z) $ is chosen such that the coefficient of $ \psi'(z) $ vanishes
\begin{equation}
	\frac{\Omega'}{\Omega} = -\frac{3}{2} \frac{a'}{a} - \frac{1}{2} \mathcal{F}
	\quad \Rightarrow \quad
	\Omega(z) = a^{-3/2}(z) \cdot f(\phi)^{-1/2}(z)\,.
\end{equation}
	Substituting into Eq.~\eqref{eq:secondorder}, the result is a Schrödinger-like equation
\begin{equation}
	- \psi''(z) + V_{\mathrm{eff}} \psi(z) = m^2 \psi(z),
	\label{schrodinger-like}
\end{equation}
where $V_{\mathrm{eff}}$ is the effective potential that is defined as

\begin{equation}
V_{\mathrm{eff}}= \frac{3}{2}\frac{a''}{a}+ \frac{3}{4}\left( \frac{a'}{a} \right)^2 +\frac{1}{2} \mathcal{F}'+\frac{3}{2} \frac{a'}{a}\mathcal{F} +\frac{1}{4} \mathcal{F}^2\,.
	\label{potential}
\end{equation}
The effective potential depends on the warp factor, the scalar field profile and the non-minimal coupling function. Equation (\ref{schrodinger-like}) can be factorized and expressed in a supersymmetric quantum mechanics form as
\begin{equation}
	L^\dagger L\psi(z) = m^2\psi(z),
\end{equation}
where
\begin{equation}
	L=\partial_z + \frac{1}{2}\left(\frac{3a'}{a} + \mathcal{F} \right)\,,
\end{equation}

and
\begin{equation}
	L^\dagger=-\partial_z + \frac{1}{2}\left(\frac{3a'}{a} + \mathcal{F} \right)\,.
\end{equation}
Note that the operator $L$ is non-negative, and the solution is stable under the TT perturbation, i.e., $m^2 \geq 0$. This ensures that there are no gravitational tachyonic modes in the system.

For the zero mode (massless mode, $m = 0$), we have $L\psi(z) = 0$, which admits the solution
\begin{equation}
	\psi_0(z) = \mathcal{N}\, a^{3/2}(z)\, f^{1/2}(\phi(z)),
\end{equation}
where $\mathcal{N}$ is the normalization constant. It is particularly noteworthy that the zero mode plays a crucial role in this framework, as it corresponds to the conventional four-dimensional massless graviton. The normalizability of this mode determines whether effective four-dimensional gravity can be localized on the brane.

The localization of the graviton zero mode is determined by the normalizability condition of its wave function in the conformal coordinate $z$, defined by $dz/dy = 1/a(y)$.  
The normalization condition can be written as
\begin{equation}
	\int_{-\infty}^{+\infty} |\psi_0(z)|^2\, dz 
	= \int_{-\infty}^{+\infty} a^{2}(y)\, f(\phi(y))\, dy < \infty,
	\label{eq:normalization}
\end{equation}
which ensures that effective four-dimensional gravity is recovered on the brane. 

\begin{figure*}[htb]
	\begin{center}
		\includegraphics[width=0.6\textwidth]{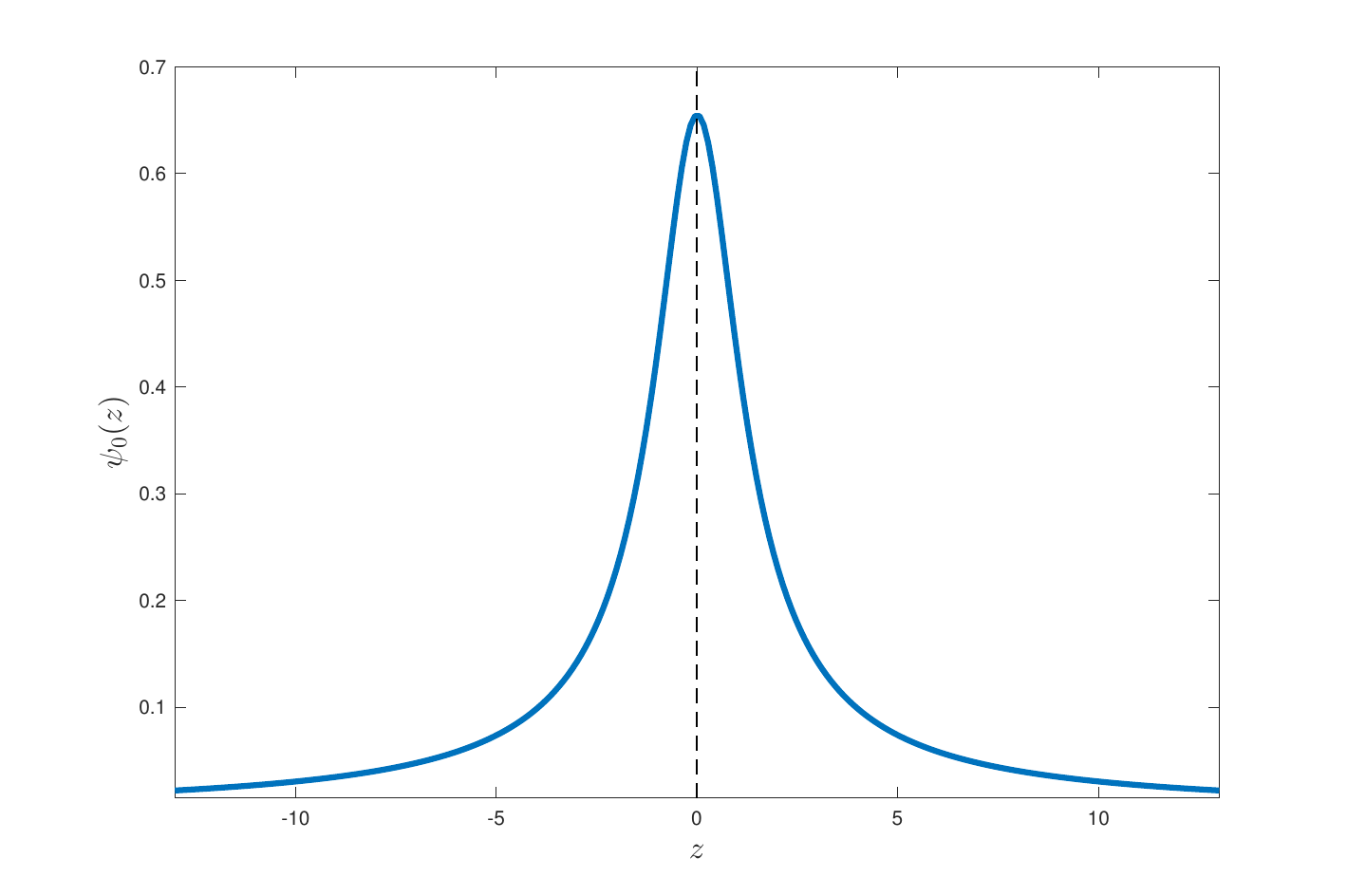} 
	\end{center}
	\caption{Plotting of gravitational zero mode $\psi_0(z)$ versus the conformal coordinate $z$ for $n=0.9$.}
	\label{zero-mode}
\end{figure*}

Since both $a(y)$ and $f(\phi)$ approach constant asymptotic values as $|y| \to \infty$, the convergence of the above integral depends critically on the power exponent of the warp factor given in Eq.~(\ref{warp}).  
For $n < 1$, this exponent is negative, causing the warp factor to decay away from the brane and ensuring that the integral in Eq.~(\ref{eq:normalization}) converges.  
In this regime, the zero mode $\psi_0(z)$ is localized around $z = 0$, confirming the recovery of four-dimensional Newtonian gravity.  
Conversely, when $n > 1$, the warp factor grows asymptotically, leading to a divergent normalization integral and hence a delocalized zero mode. Therefore, the parameter $n$ governs the localization behavior of gravity, and this analytical result is in full agreement with the numerical profiles of the zero mode presented in Fig.~\ref{zero-mode}.
Our numerical analysis further confirms that the zero mode $\psi_0(z)$ is normalizable and sharply peaked around $z = 0$, indicating that effective four-dimensional gravity is successfully localized on the brane.

\begin{figure*}[htb]
	\begin{center}
		\includegraphics[width=0.6\textwidth]{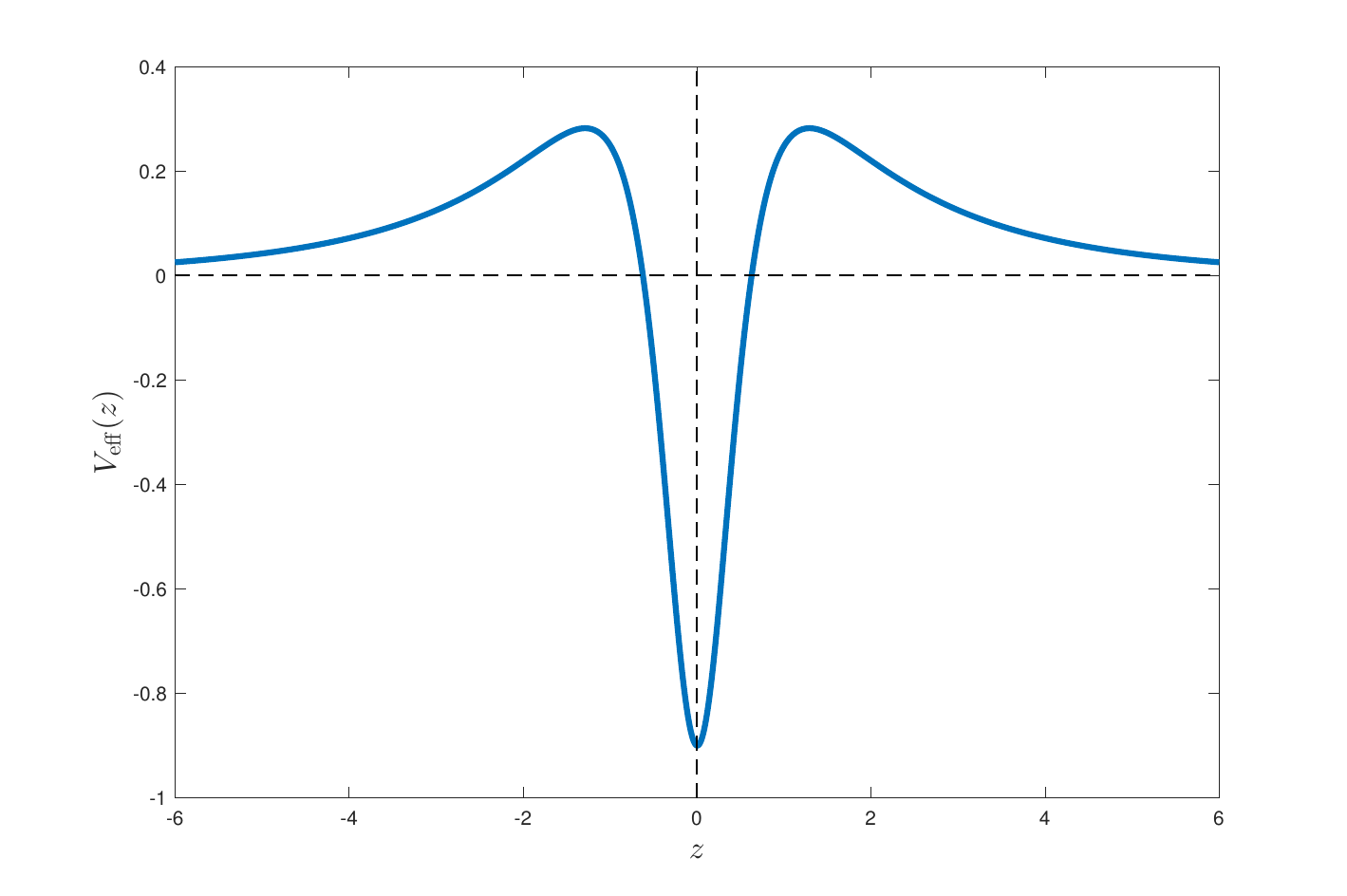} 
	\end{center}
	\caption{The effective potential $V_{\mathrm{eff}}(z)$ for tensor perturbations in the conformal coordinate $z$, corresponding to a thick brane model.}	\label{fig:Veff_z}
\end{figure*}

The effective potential $V_{\mathrm{eff}}(z)$ derived from the Schrödinger-like equation associated with tensor perturbations is depicted in figure \ref{fig:Veff_z} . As the figure shows, the effective potential exhibits a characteristic volcano-like shape. It is symmetric around $ z = 0 $, where it reaches a minimum. This central dip in the potential plays a key role in trapping the graviton zero mode on the brane. The existence of such a normalizable bound state ensures that four-dimensional gravity is recovered on the brane at low energies.
 
\begin{figure*}[htb]
	\begin{center}
		\includegraphics[width=0.6\textwidth]{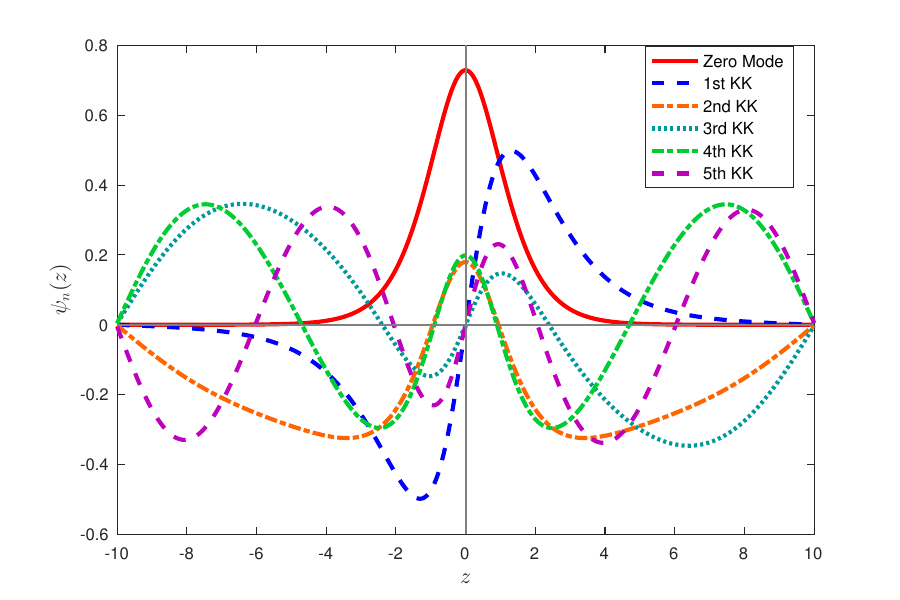} 
	\end{center}
	\caption{Profiles of the zero mode and the first five Kaluza-Klein (KK) modes $\psi_n(z)$ of a bulk field in a thick brane scenario. The zero mode (red solid line) is sharply localized at the center of the brane ($z = 0$), ensuring recovery of standard 4D physics. Higher KK modes correspond to massive excitations with increasing numbers of nodes, becoming progressively more delocalized in the bulk. }	\label{fig:kk}
\end{figure*}

The volcano-like potential structure implies the existence of a continuous spectrum of Kaluza-Klein excitations above the zero mode. These massive modes have wavefunctions that are only weakly localized near the brane, may be a result of small corrections to Newtonian gravity at sub-millimeter distances.
Figure~\ref{fig:kk} presents the wavefunction profiles $\psi_n(z)$ of the zero mode and the first five KK excitations of a bulk field along the extra-dimensional conformal coordinate $z$, in the framework of our thick brane model. The \textbf{first to fifth KK modes} (depicted with distinct dashed and dotted lines) represent the \textit{massive modes} of the bulk field. These modes are \textit{less localized} and exhibit increasing \textit{oscillatory behavior} with higher mode numbers:
\begin{itemize}
	\item The \textbf{1st KK mode} has one node,
	\item The \textbf{2nd KK mode} has two nodes, and so on.
\end{itemize}
Their spatial profiles extend further into the bulk, indicating that \textbf{massive KK states are quasi-localized or delocalized}, depending on the shape of the effective potential in the Schrödinger-like equation governing the mode functions.

\subsection*{Phenomenological Implications}

The existence of a normalizable graviton zero mode ensures that standard four-dimensional gravity is recovered on the brane at low energies. 
In this case, the effective four-dimensional Planck mass is finite and proportional to the normalization integral of the zero mode. 
The massive Kaluza-Klein (KK) excitations correspond to higher-dimensional gravitational fluctuations that can propagate into the bulk and induce small corrections to the Newtonian potential on the brane. 
These corrections typically take the form
\begin{equation}
	V(r) \simeq \frac{G_N m_1 m_2}{r} 
	\left[ 1 + \int_{m_{\text{gap}}}^{\infty} dm\, \rho(m)\, e^{-mr} \right],
\end{equation}
where $\rho(m)$ denotes the spectral density of KK states and $m_{\text{gap}}$ represents the possible mass gap between the zero mode and the continuum spectrum.

In the present model, the effective potential $V_{\mathrm{eff}}(z)$ exhibits a volcano-type profile that supports a quasi-localized zero mode and a continuous KK spectrum with a small or vanishing mass gap. 
As a result, the leading-order corrections to Newton’s law are exponentially suppressed at distances larger than the brane thickness, remaining consistent with current submillimeter tests of gravity. 
At higher energies, however, the excitation of massive KK modes may affect the propagation of gravitational waves in the bulk, producing potential dispersion or damping effects. 
Such deviations could, in principle, be probed by present and future gravitational-wave observatories, including LIGO-Virgo and LISA, offering an observational window into the extra-dimensional structure of spacetime.

A detailed study of the gravitational-wave phenomenology, including possible constraints on the parameters $(\xi, n, b)$ from observational data, is left for future investigation.

\section{Discussions and Conclusion}\label{sec4}

In this work, we explored a thick brane model within the Palatini formulation of gravity, where the metric and affine connection are treated as independent variables. By introducing a non-minimal coupling between a bulk scalar field and the Ricci scalar, we derived the modified field equations and obtained analytic solutions under the assumption of a flat metric with four-dimensional Poincaré symmetry and a kink-like scalar field configuration.

The resulting warp factor exhibits a bell-shaped profile centered at $y=0$, while the scalar potential $V(y)$ displays a symmetric volcano-like structure with a central minimum, characteristic of thick brane scenarios. The energy density $\rho(y)$ shows a central peak, two symmetric negative minima, and vanishes asymptotically, all consistent with a regular, finite-thickness brane. Additionally, the scalar potential $V(\phi)$ has a periodic, oscillatory structure with a growing envelope, ensuring scalar field confinement and contributing to the stability of the solution.

We also investigated the stability of the background solution and the localization of gravity through linear tensor perturbations. The corresponding Schrödinger-like equation, derived from the Kaluza--Klein decomposition, admits a supersymmetric factorization that guarantees the absence of tachyonic modes. The effective potential exhibits a symmetric, volcano-like shape with a central minimum at $z=0$, allowing the graviton zero mode to be localized on the brane. 
Our analytical and numerical analyses consistently confirm that the graviton zero mode $ \psi_0(z) $ satisfies the normalization condition and is sharply peaked around the brane location. This ensures the localization of four-dimensional gravity and validates the internal consistency of the model within the Palatini framework. 
Furthermore, we computed the first few massive Kaluza--Klein modes, which represent higher-energy excitations with increasing numbers of nodes and are progressively delocalized in the bulk.

It is instructive to compare our findings with previous investigations of thick branes involving non-minimal couplings. In the metric formalism, Guo \emph{et al.}~\cite{Guo:2011wr} obtained analytic kink-like configurations and analyzed tensor perturbations, demonstrating gravity localization. In contrast, our study is developed within the Palatini formulation, where the metric and connection are treated independently. This difference leads to modified field equations and a distinct effective potential structure, altering the localization mechanism of the graviton zero mode. Additionally, Bazeia \emph{et al.}~\cite{Bazeia:2014poa} constructed thick-brane solutions in Palatini $ f(R) $ gravity, focusing on the warp factor, scalar potential, and energy density. However, their work did not include a stability analysis or Kaluza--Klein decomposition. Our results therefore extend the Palatini braneworld framework by providing a complete tensor perturbation analysis, computing the KK spectrum, and explicitly demonstrating the stability and gravitational localization of the model. We also emphasize that our numerical treatment of the Schrödinger-like equation provides explicit profiles for the first few KK modes. To the best of our knowledge, such explicit mode visualizations are rarely presented in Palatini-based thick brane models, and thus represent a methodological improvement and a step toward connecting the theoretical framework with possible phenomenological implications.

Looking forward, this framework can be extended in several directions. One natural continuation is to examine the localization of fermionic and scalar fields within the same Palatini setting, which would provide a more comprehensive understanding of matter confinement. In particular, fermion localization typically requires introducing a Yukawa-type coupling between the bulk fermion and the background scalar field. Such an interaction can generate an effective potential capable of trapping chiral fermionic zero modes on the brane. It would also be interesting to explore generalized coupling functions or more intricate scalar potentials, which may give rise to double-kink or multi-kink structures with richer phenomenology. Finally, connecting the properties of the localized graviton zero mode and the massive KK spectrum to potential gravitational-wave observations could offer a pathway to testing Palatini braneworld scenarios in future experiments.

We also note that, in a recent related work~\cite{alimoradi2025}, we examined a thin Palatini braneworld model with a delta-function brane, bulk cosmological constant, and brane tension. That study focused on cosmological implications and the modified Friedmann dynamics arising from the non-minimal coupling between the bulk scalar field and the Palatini Ricci scalar. Although the setup differs from the present thick-brane configuration without brane matter, both analyses complement each other in understanding how the Palatini formulation affects gravitational and cosmological behavior in braneworld scenarios.

Overall, our results demonstrate that Palatini gravity with a non-minimally coupled scalar field provides a viable and self-consistent framework for constructing stable thick brane models. The analysis presented here clarifies the distinctive features of the Palatini approach and highlights its potential relevance to gravity localization and braneworld phenomenology in modified theories of gravity.

\bibliographystyle{unsrt}
\bibliography{citation-brane}
%
%
%

%
%
	
\end{document}